\documentclass[pra,twocolumn,epsfig,rotate,superscriptaddress,showpacs]{revtex4}
\usepackage{graphicx}
\usepackage{epstopdf}
\usepackage{amsfonts}
\usepackage{amssymb}
\usepackage{amsmath}
\usepackage{subfigure}
\usepackage{prettyref}
\usepackage{float}
\usepackage{color,CJK}
\usepackage{amsmath}
\usepackage{amssymb}
\usepackage{esint}

\usepackage{braket}

\def\be{\begin{equation}}
\def\ee{\end{equation}}
\def\ba{\begin{eqnarray}}
\def\ea{\end{eqnarray}}

\begin{document}
\begin{CJK}{UTF8}{gbsn}
\title
{Non-Hermitian quantum systems and their geometric phases}

\author{Qi Zhang(张起)}
\affiliation{College of Science, Zhejiang University of Technology,
Hangzhou 310023, China}
\author{Biao Wu(吴飙)}
\affiliation{International Center for Quantum Materials, Peking University, 100871, Beijing, China}
\affiliation{Wilczek Quantum Center, School of Physics and Astronomy, Shanghai Jiao Tong University, Shanghai 200240, China}
\date{\today}
\begin{abstract}
We discuss the basic theoretical framework for non-Hermitian quantum systems with particular
emphasis on the diagonalizability of  non-Hermitian Hamiltonians and
their $GL(1,\mathbb{C})$ gauge freedom, which are relevant to  the adiabatic evolution
of non-Hermitian quantum systems.
We find that the adiabatic evolution is possible only when the eigen-energies are real.
The accompanying geometric phase is found to be generally complex and associated with
not only the phase of a wavefunction but also its amplitude.
The condition for the real geometric phase is laid out. Our results are illustrated with two examples
of non-Hermitian $\mathcal{PT}$ symmetric systems,
the two-dimensional non-Hermitian Dirac fermion model and bosonic Bogoliubov quasi-particles.
\end{abstract}

\maketitle
\end{CJK}

\section{Introduction}
Nature is fundamentally described by the familiar quantum mechanics, where
Hamiltonians and observables are all Hermitian operators. Nevertheless,
due to approximation or interaction with environment, non-Hermitian quantum systems
do arise, for example, bosonic Bogoliubov systems~\cite{ZhangNJP,njp} and non-Hermitian $\mathcal{PT}$ symmetric
systems~\cite{Bender}. There have been tremendous interests recently in these non-Hermitian systems
both theoretically ~\cite{Bender,Bender2,Bender3,Mostafazadeh,Berry,Longhi,West,Mostafazadeh2,Bender4,Bender5,Mostafazadeh3,Mostafazadeh4,Wang,chen} and experimentally~\cite{e1,e2,e3,e4,e5,e6,e7,e8,e9,e10,e11,e12,e13,J1,J3,J4,J5}.

There has also been a growing interest in topological properties of non-Hermitian Hamiltonians~\cite{T1,T2,T3,T4,T5,T6,T7,T8,T9,Fu,chen1,chen2},
where Chern number associated with Berry curvature is introduced to characterize the topology  and the existence of edge states.
However, many basic issues are yet to be clarified. For instance, in Hermitian systems, Berry phase
is defined only when the adiabatic evolution is possible. It is not clear what ensures the adiabatic evolution
in non-Hermitian systems.

In this work we investigate the adiabatic evolution and its associated geometric phase in non-Hermitian systems.
To set up the theoretical framework for discussion, we first describe basic features of non-Hermitian systems. They
include the diagonalizability of a non-Hermitian Hamiltonian, $GL(1,\mathbb{C})$ gauge transformation, and non-orthonormal basis imposed
on the Hilbert space by a non-Hermitian Hamiltonian. Due to the last feature, one vector in the Hilbert space has two different forms,
covariant and contravariant. We find that the adiabatic evolution is possible  only when the eigen-energies are real.
In general, non-Hermitian systems have a  gauge freedom of $GL(1,\mathbb{C})$,
and therefore  the geometric phase is generally complex and associated with both phase and amplitude of the eigenstate.
However, the geometric phase can be real when certain conditions are satisfied.
Our results are illustrated with two  two-mode non-Hermitian systems, the two-dimensional non-Hermitian
Dirac fermion model and bosonic Bogoliubov quasi-particles..

\section{General features of  non-Hermitian systems}
Non-Hermitian systems  share many basic features with the usual Hermitian systems. For example,
their states live in Hilbert spaces and all observables except energy
are represented by Hermitian operators. At the same time, non-Hermicity  brings new features.
We discuss the features that are relevant to adiabatic evolution and geometric phase.
The first is the diagonalizability of a non-Hermitian Hamiltonian, which is related to the exceptional points (EPs) in a parameter space.
The second is that the non-Hermitian Hamiltonian imposes two sets of non-orthonormal basis, which are biorthonormal to each other,
in the Hilbert space. We find  it very natural to  use covariant vector and contravariant vector to deal with this issue.
The third is the gauge freedom in a non-Hermitian system. The norm is not conserved in a non-Hermitian system,
thus its gauge freedom is of  $GL(1,\mathbb{C})$ in contrast to $U(1)$ gauge freedom in a Hermitian system.
As a result, geometric phases are in general complex. We find that conserved psudo-norms can be defined
when eigenvalues of the non-Hermitian Hamiltonian are real. Furthermore, geometric phases may become real when more
conditions are satisfied.

\subsection{Diagonalizability of non-Hermitian Hamiltonians}
We consider a general $n$-dimensional matrix ${\cal M}$.  Its  diagonalizablity is determined by
the algebraic multiplicity and geometric multiplicity of its
eigenvalues. The eigenvalues $\lambda_1, \lambda_2,\cdots,\lambda_d$ of  matrix ${\cal M}$
are the roots of the secular equation,
\begin{eqnarray} \label{alge}
&&\text{Det}({\cal M}-\lambda I)\\
&=&(\lambda_1-\lambda)^{\eta(\lambda_1)}(\lambda_2-\lambda)^{\eta(\lambda_2)}\ldots(\lambda_d-\lambda)^{\eta(\lambda_d)}=0,\nonumber
\end{eqnarray}
where $I$ is the identity matrix and  $\text{Det}(A)$ denotes the determinant of $A$.
In the polynomial the exponent $\eta(\lambda_j)$ is called
algebraic multiplicity of eigenvalue $\lambda_j$~\cite{Nering}.
The following relations apparently hold  for the algebraic multiplicity,
\begin{equation}
1\leq \eta(\lambda_j)\leq n, \quad \sum_{i=1}^d \eta(\lambda_i)=n.
\end{equation}

Corresponding to each eigenvalue $\lambda_j$, the maximum number of linearly independent eigenvectors  is called  geometric
multiplicity $\zeta(\lambda_j)$~\cite{Nering}.
One can prove that for each eigenvalue $\lambda_j$ its geometric multiplicity cannot exceed its algebraic multiplicity,
that is,  $\zeta(\lambda_j)\leq\eta(\lambda_j)$~\cite{Nering}.
The matrix ${\cal M}$ is diagonalizable only when the geometric multiplicity is equal to the algebraic multiplicity for any eigenvalue~\cite{Nering},
\begin{equation} \label{DI}
\zeta(\lambda_j)=\eta(\lambda_j), ~~~~\; \text{for} \; 1\leq j\leq d.
\end{equation}

Consider a family of non-Hermitian Hamiltonians $H(\mathbf{R})\neq H(\mathbf{R})^\dag$, which depend on external parameters $\mathbf{R}$.
The points in the parameter space $\mathbf{R}$ are called exceptional points (EPs) when $H(\mathbf{R})$ is not  diagonalizable at these points.
For  parameters other than EPs, Eq.~(\ref{DI}) holds and there are $n$ linearly independent eigenvectors for an $n\times n$ Hamiltonian matrix $H$.

\subsection{Covariant and contravariant vectors  in Hilbert space}
When a non-Hermitian Hamiltonian $H$ is diagonalizable, it has two sets of eigenvectors
$|\psi_j\rangle$ and $|\phi^j\rangle$ satisfying~\cite{F3,F1,F2}
\be
H|\psi_j\rangle=E_j|\psi_j\rangle,  \quad
H^\dag|\phi^j\rangle=E_j^*|\phi^j\rangle,. \\  \label{right-left}
\ee
They are  biorthonormal, $\langle\phi^i|\psi_j\rangle=\delta_{ij}$, and complete
\be  \label{right-left2}
\sum_j |\psi_j\rangle\langle\phi^j|=1\,.
\ee
Usually $|\psi_j\rangle$ and $|\phi^j\rangle$ are called right eigenvector and  left eigenvector.  We find it more natural to call them
contravariant eigenvectors and covariant eigenvectors. Respectively, they form one set of contravariant basis and one set of covariant basis.
For a given vector $\ket{\Psi}$ in the Hilbert space, it can be expanded either in the contravariant basis
\begin{equation}
|\Psi\rangle=\sum_{j=1}^n c^j|\psi_j\rangle\equiv\left(\begin{array}{c} c^1\\c^2\\\vdots\\c^n \end{array}\right),
\end{equation}
or   in the covariant basis
\begin{equation}
|\Psi\rangle=\sum_{j=1}^n c_j|\phi^j\rangle\equiv\left(\begin{array}{c} c_1\\c_2\\\vdots\\c_n \end{array}\right)\,.
\end{equation}
The inner product can be naturally written as
\be
\braket{\Psi|\Psi}=\sum_{j=1}^nc_j^*c^j\,.
\ee
Note that in the above  we have introduced  upper and lower indices to label  covariant and contravariant vectors, respectively.

\subsection{Gauge freedom and psudo-norms in non-Hermitian systems}
Consider the dynamics of a non-Hermitian system, which is given by the Schr\"odinger equation
\begin{equation}
\label{Schro}
\text{i}\hbar\frac{\partial}{\partial t}|\Psi\rangle=H|\Psi\rangle\,.
\end{equation}
As $H$ is not Hermitian, the norm $\braket{\Psi|\Psi}$ is not conserved during the dynamical evolution.
This means that we can carry out a transformation of the wavefunction $\ket{\Psi'}=f\ket{\Psi}$
with $f=|f|e^{\text{i}\theta}$ and $|f|\neq1$. This is  a general linear (GL) gauge transformation in complex domain.
Therefore, in general, a non-Hermitian system has $GL(1,\mathbb{C})$ gauge freedom.

However, as we shall show immediately, for a class of non-Hermitian systems, one can define a  psudo-norm that is conserved.
 For the Hilbert space,  there always exists a set of complete orthonormal basis $|j \rangle$, $\langle i|j\rangle=\delta_{ij}$.
When $H$ is diagonalizable,  although the right eigen-vectors $\ket{\psi_j}$ are not orthonormal  they are linearly independent
and form a set of complete basis. The same is true for the left  eigen-vectors $\ket{\phi^j}$. Therefore, there exists a convertible
matrix $A$ such that
\be \label{relation}
|\psi_j \rangle = A | j\rangle\,,~~~ |\phi^j\rangle = (A^{-1})^\dag|j\rangle,
\ee
 Thus we have $|\phi^j\rangle=(A^{-1})^\dag A^{-1}|\psi_j\rangle$ and,
\begin{equation} \label{inva1}
\langle\psi_i |X|\psi_j \rangle=\braket{\psi_i|\phi^j}=\delta_{ij},
\end{equation}
where $X(\mathbf{R})=(AA^\dag)^{-1}$ and is apparently Hermitian. We define the psudo-norm as
$\langle\psi |X|\psi\rangle$. One can easily prove that
\be
\frac{d}{dt}\langle\psi |X|\psi\rangle=0\,
\ee
when all the eigenvalues $E_j$'s are real. A special  case of such a norm is well known in Bogoliubov systems~\cite{njp}.
We will find later that Hermitian matrix $X$ plays a crucial role in specifying the condition for geometric phase to be real.

\section{Adiabatic Evolution}
Consider a non-Hermitian Hamiltonian $H(\mathbf{R})$, which depends on external parameters $\mathbf{R}$.
We are interested in its adiabatic evolution  as $\mathbf{R}$ changes slowly with time
and how geometric phase arises. In conventional quantum mechanics where $H(\mathbf{R})$ is Hermitian,
there is an adiabatic theorem which states that the occupation probability at each energy level does not change
when there is no degeneracy in energy levels.
Berry later found that geometric phase can arise when the adiabatic theorem holds.
We want to find out under what condition a similar adiabatic theorem holds in non-Hermitian systems.

We first assume that $\mathbf{R}$ is fixed. In this case, as the system is linear, we can
always expand a state $|\Psi(t)\rangle$ in terms of the right eigenstates
and write the dynamical evolution as
\begin{equation} \label{super}
|\Psi(t)\rangle=\sum_j c^j \exp[-\frac{\text{i}}{\hbar}E_j t]|\psi_j\rangle\,.
\end{equation}
This shows that if $E_j$'s are complex then the relative probability in each eigenstate $|\psi_j\rangle$ can change with time.
The situation may become worse when $\mathbf{R}$ changes. So, for an adiabatic theorem to hold
in non-Hermitian systems, the eigenvalues $E_j$'s must be real and have no degeneracy.
This conclusion becomes more evident with the following detailed analysis.

When $\mathbf{R}$ changes with time, all the eigenstates $|\psi_j(t)\rangle$ and eigenvalues $E_j(t)$
become time dependent. In this case, we can write the dynamical evolution as
\begin{equation} \label{super}
|\psi(t)\rangle=\sum_j c^j(t) \exp[-\frac{\text{i}}{\hbar}\int_0^tE_j(t')dt']|\psi_j(t)\rangle.
\end{equation}
We substitute it into the Schr\"odinger equation (\ref{Schro}) and obtain using Eq. (\ref{right-left})
\begin{eqnarray} \nonumber  \label{6}
&\text{i}\hbar\sum_j\dot{c}^j(t)\exp[-\frac{\text{i}}{\hbar}\int_0^tE_n(t')dt']|\psi_j(t)\rangle\\
&+\text{i}\hbar\sum_jc^j(t)\exp[-\frac{\text{i}}{\hbar}\int_0^tE_j(t')dt']|\dot{\psi}_j(t)\rangle=0.
\end{eqnarray}
Multiplying Eq.~(\ref{6}) with the left eigenstate $\langle \phi^m(t)|$, we have
\begin{eqnarray} \label{7}
&&\dot{c}^m=-c^m\langle\phi^m|\dot{\psi}_m\rangle \\   \nonumber
&&-\sum_{j\neq m}c^j\langle\phi^m|\dot{\psi}_j\rangle \exp\left[-\frac{\text{i}}{\hbar}\int_0^t(E_j(t')-E_m(t'))dt'\right].
\end{eqnarray}
We assume that the system is initially in  state  $|\psi_m\rangle$. If the adiabatic theorem holds, one would have
$c^m\sim 1$ and $|c^j|\ll 1$ $(j\neq m)$ during the whole process. When $E_j$'s are all real, the second term on the right hand
side of the above equation can be safely neglected as,
\begin{equation} \label{condi}
\left| \frac{\hbar\langle\phi^m|\dot{\psi}_j\rangle}{E_m-E_j} \right|\ll 1, \text{for all} \; j\neq m.
\end{equation}
This can be found by integrating Eq.~(\ref{7}). We then have
\be \label{adiasolution}
\dot{c}^m=-c^m\langle\phi^m|\dot{\psi}_m\rangle.
\ee
This is similar to the situation in Hermitian systems. When $E_j$'s are complex, the second term can grow exponentially
and can not be neglected. This means  that  the adiabatic theorem can not hold when $E_j$'s are complex.
From now on we only consider the real eigenvalue case.

\section{Geometric phase}
We are now ready to derive geometric phase. We assume that the system in state $\ket{\psi_j}$. When the adiabatic theorem
holds, it should evolve with time as
\begin{equation} \label{e26}
|\psi(\mathbf{R})\rangle=|\psi_j(\mathbf{R})\rangle e^{-{\text i}\frac{\int E_j(\mathbf{R}) dt}{\hbar}}e^{{\text i}\beta_j}\,,
\end{equation}
where  $\beta_j$ is the  geometric phase. According to Eq. (\ref{adiasolution}), we have ~\cite{GarrisonPLA,GePRA},
\begin{equation} \label{Berry-conn}
\mathbf{A}_j=\frac{\partial \beta_j}{\partial \mathbf{R}}=\text{i}\langle \phi^j(\mathbf{R})|\frac{\partial}{\partial \mathbf{R}}|\psi_j(\mathbf{R})\rangle.
\end{equation}
The Berry curvature  therefore takes the following form~\cite{GarrisonPLA},
\begin{equation} \label{Gcur}
\mathbf{B}_j=\text{i}\langle\nabla\phi^j|\times|\nabla\psi_j\rangle,
\end{equation}
where $\nabla\equiv\frac{\partial}{\partial\mathbf{R}}$. Because $|\psi_j\rangle$ is usually not equal to $|\phi^j\rangle$ (as $H^\dag\neq H$), the Berry connection and Berry curvature are generally not real for non-Hermitian systems even when all $E_j$'s are real.

Let us now examine under what condition  the Berry connection in (\ref{Berry-conn}) is real. As we are
considering the case where all $E_j$'s are real,  Eq.~(\ref{inva1}) holds.
Differentiating Eq.~(\ref{inva1}) with respect to $\mathbf{R}$,
we get
\begin{equation} \label{e32}
\langle \psi_j|X\frac{\partial }{\partial \mathbf{R}}|\psi_j\rangle+\left(\langle \psi_j|X\frac{\partial }{\partial \mathbf{R}}|\psi_j\rangle\right)^*+\langle \psi_j|\frac{\partial X}{\partial \mathbf{R}}|\psi_j\rangle=0.
\end{equation}
where we have taken advantage of $X$ being Hermitian.
Therefore, when $X$ is $\mathbf{R}$-dependent,  the following quantity is in general not zero,
\begin{equation} \label{e33}
\langle\psi_j|X\frac{\partial }{\partial \mathbf{R}}|\psi_j\rangle+\left(\langle \psi_j|X\frac{\partial }
{\partial \mathbf{R}}|\psi_j\rangle\right)^*\neq0\,.
\end{equation}
This implies that  $\langle \psi_j|X\frac{\partial }{\partial \mathbf{R}}|\psi_j\rangle$ may not be purely imaginary
and thus $\mathbf{A}_j$ may not be real.   The Berry connection is real only if the following identity holds
\begin{equation} \label{conditionY}
\langle \psi_j|\frac{\partial X}{\partial \mathbf{R}}|\psi_j\rangle=0\,.
\end{equation}
It is  important to note that the above condition is {\it not} equivalent to $\frac{\partial X}{\partial \mathbf{R}}=0$.
It is possible that the above condition holds when $\frac{\partial X}{\partial \mathbf{R}}\neq 0$. The reason is that
the matrix $X$ is independent of choices of $\ket{j}$ and is completely determined by $\ket{\psi_j}$. Another way to understand this
is to note that the condition (\ref{conditionY}) is {\it not} equivalent to
\be
\langle \Psi|\frac{\partial X}{\partial \mathbf{R}}|\Psi\rangle=0\,,
\ee
where $\ket{\Psi}$ is an arbitrary vector in the Hilbert space.

Nevertheless, we find that for many non-Hermitian systems where Eq.~(\ref{conditionY}) holds we can find an $\mathbf{R}$-independent Hermitian matrix $Y$ such that
\begin{equation} \label{relation25}
|\phi^j(\mathbf{R})\rangle=\alpha_j Y|\psi_j(\mathbf{R})\rangle=X(\mathbf{R})|\psi_j(\mathbf{R})\rangle, \;{\text{for}} \; j=1,2,\ldots,
\end{equation}
with $\alpha_j=\pm1$.
Differentiating Eq.~(\ref{relation25}) with respect to $\mathbf{R}$ and left multiplying it with $\langle\phi^j|$, we can still obtain the relation (\ref{conditionY}) by virtue of Hermiticity of $X$ and $Y$ and $dY/d\mathbf{R}=0$. With the constant $Y$, the following relation holds
\begin{equation} \label{relation21}
\langle\psi_j(\mathbf{R})|Y|\psi_j(\mathbf{R})\rangle=\alpha_j,
\end{equation}
and the Berry connection can be expressed as,
\begin{equation} \label{Berry-conn2}
\mathbf{A}_j=\alpha_j\text{i}\langle \psi_j(\mathbf{R})|Y|\frac{\partial}{\partial \mathbf{R}}\psi_j(\mathbf{R})\rangle\,.
\end{equation}

According to the gauge freedom in non-Hermitian system considered in Sec.~IIC, 
there is a freedom to modify the $j$th eigenstate by a complex number $f$ ($|f|\neq1$ and $f\in GL(1,\mathbb{C})$),
\begin{equation} \label{gauge55}
|\psi_j'\rangle=f|\psi_j\rangle, \; \langle\phi^{'j}|=\frac{1}{f}\langle\phi^j|.
\end{equation}
The second equation in (\ref{gauge55}) is to guarantee the biorthonormal condition.
Upon the gauge transformation (\ref{gauge55}) the Berry connection is modified to,
\begin{equation}  \nonumber
\mathbf{A}_j'=\mathbf{A}_j+\text{i}\frac{1}{f}\frac{\partial f}{\partial \mathbf{R}}.
\end{equation}
Writing $f=|f|e^{\text{i}\theta}$ we have,
\begin{equation} \label{gaugetrans}
\mathbf{A}_j'=\mathbf{A}_j+\text{i}\frac{1}{|f|}\frac{\partial |f|}{\partial \mathbf{R}}-\frac{\partial \theta}{\partial \mathbf{R}},
\end{equation}
When $|f|=1$ we recover the result for Hermitian systems. Furthermore, it can be checked that
\begin{equation}
|\psi_j'\rangle e^{\text{i}\int_{R_1}^{R_2}\mathbf{A}_j'd\mathbf{R}}=|\psi_j\rangle e^{\text{i}\int_{R_1}^{R_2}\mathbf{A}_jd\mathbf{R}}\,,
\end{equation}
indicating that the current framework for the geometric phase is self-contained.

\section{Monopoles}
We have defined exceptional points (EPs) as points in the parameter space
$\mathbf{R}$ where non-Hermitian matrix $H(\mathbf{R})$ is not diagonalizable.
In this section we shall show that they are monopoles in the sense that
the divergence of the Berry curvature $\nabla\cdot\mathbf{B}_j$ does not vanish.

According to Eq.~(\ref{Gcur}), the Berry curvature can be written as,
\begin{eqnarray}
\mathbf{B}_j=\nabla\times \mathbf{A}_j
={\text i}\sum_{j'} \langle\nabla \phi^j|\psi_{j'}\rangle\times\langle \phi_{j'}|\nabla \psi_j\rangle\,, \label{B1}
\end{eqnarray}
where  the completeness condition in (\ref{right-left2}) is employed.
To  calculate the divergence of the Berry curvature, i.e. $\nabla\cdot\mathbf{B}_j$, we introduce an auxiliary operator
\begin{equation} \label{e37}
\mathbf{F}=-{\text i}\sum_{n}|\nabla \psi_{n}\rangle\langle \phi_{n}|={\text i}\sum_{n}|\psi_{n}\rangle\langle\nabla\phi_{n}|,
\end{equation}
where the second equality is ensured by the completeness relation (\ref{right-left2}). It can be checked that
\be \label{F-re}
\nabla\times\mathbf{F}=-{\text i}\mathbf{F}\times \mathbf{F}\,.
\ee
The Berry curvature can be expressed in terms of $\mathbf{F}$ as
\begin{equation}
\mathbf{B}_j={\text i} \sum_{j'} \langle \phi^j|\mathbf{F}|\psi_{j'}\rangle\times\langle \phi_{j'}|\mathbf{F}|\psi_j\rangle
={\text i}\langle \phi^j|\mathbf{F}\times \mathbf{F}|\psi_j\rangle.
\end{equation}
Finally, by virtue of  Eq.~(\ref{F-re}), we find
\begin{eqnarray} \label{e41}  \nonumber
&&\nabla\cdot\mathbf{B}_j\\ \nonumber
&=&{\text i}[\langle\nabla \phi^j|\cdot(\mathbf{F}\times \mathbf{F})|\psi_j\rangle+
\langle \phi^j|(\mathbf{F}\times \mathbf{F})\cdot|\nabla\psi_j\rangle \\ \nonumber
&&+\langle \phi^j|\nabla\cdot(\mathbf{F}\times\mathbf{F})|\psi_j\rangle] \\ \nonumber
&=&{\text i}[-{\text i}\langle \phi^j|\mathbf{F}\cdot(\mathbf{F}\times \mathbf{F})|\psi_j\rangle
+{\text i}\langle \phi^j|(\mathbf{F}\times\mathbf{F})\cdot \mathbf{F}|\psi_j\rangle \\ \nonumber
&&+\langle \phi^j|(\nabla\times\mathbf{F})\cdot \mathbf{F}|\psi_j\rangle-\langle \psi_j|\mathbf{F} \cdot(\nabla\times\mathbf{F})|\psi_j\rangle] \\
&=&0.
\end{eqnarray}
In the above derivation we have used the completeness relation (\ref{right-left2}), which is equivalent to that
$H(\mathbf{R})$ is diagonalizable. Therefore, for all the points in the parameter space
$\mathbf{R}$ other than EPs, the divergence of the Berry curvature is zero.  In other words, monopoles can only be EPs.

\section{Examples}
In the above we have presented a general framework for geometric phases in non-Hermitian systems.
In this section we use two simple examples to illustrate these results. Specifically, these two examples
are a Dirac model with non-Hermitian terms and a two-mode Bogoliubov de Gennes model describing the Bosonic Bogoliubov quasiparticles. \\

\subsection{non-Hermitian Dirac model}

As the first illustrative example, we investigate the Dirac model with a non-Hermitian term.
The Hamiltonian is
\begin{equation} \label{gene-H}
H=p_x\sigma_x+p_y\sigma_y+(p_z+\text{i}s)\sigma_z\,,
\end{equation}
where  $p_x$, $p_y$ and $p_z$ are  the Bloch momentum and $s$ is a real constant, denoting the gain and loss of particles.
$\sigma_x$, $\sigma_y$ and $\sigma_z$ are Pauli matrices. This non-Hermitian Dirac model has recently studied
to reveal the topology of energy bands and the properties of edge state~\cite{T9,Fu,chen1,chen2}.
The energy bands of $H$ are
\begin{equation} \label{Gen}
E_{1(2)}=\pm\sqrt{\mathbf{p}^2-s^2+2\text{i}(p_zs)}\,.
\end{equation}
They are real when  $p_z=0$ and $p_x^2+p_y^2\geq s^2$. In particular, $E_1=E_2=0$
on the ring $p_x^2+p_y^2=s^2$ at $p_z=0$.  As we shall show that this ring is a collection of EPs,
where $H$ becomes non-diagonalizable. It is worth noting that we found a disk-shaped monopole
in a nonlinear quantum system~\cite{PLA}.

For a point on the ring $p_x^2+p_y^2=s^2$ at $p_z=0$, we can obtain the algebraic multiplicity $\eta(0)$
according to Eq.~(\ref{alge}), and the geometric multiplicity $\zeta(0)$ by  examining the number of  linearly
independent eigenstates with zero eigenvalue. The result is $\eta(0)=2$ and $\zeta(0)=1$, violating the diagonalizable condition (\ref{DI}).
This means that all points on the ring are EPs.  For any point off the ring, we have $\eta(E_1)=\zeta(E_1)=\eta(E_2)=\zeta(E_2)=1$,
which satisfies the diagonalizable condition. In other words, all EPs are on the ring $p_x^2+p_y^2=s^2$ at $p_z=0$.

When $H$ in Eq. ((\ref{gene-H})) is diagonalizable, its biorthonormal eigenstates
corresponding to the eigenenergy $E_1$ and $E_2$ are
\begin{eqnarray}  \label{wf1} \nonumber
|\psi_1\rangle&=&\left(\begin{array}{c} \sqrt{\mathbf{p}^2-s^2+2\text{i}p_zs}+is+p_z \\ \\ p_x+\text{i}p_y  \end{array}\right) \\
|\phi^1\rangle&=&\left(\begin{array}{c} \frac{1}{2\sqrt{\mathbf{p}^2-s^2-2\text{i}p_zs}} \\ \\
\frac{\sqrt{\mathbf{p}^2-s^2-2\text{i}p_zs}+\text{i}s-p_z}{2(p_x-\text{i}p_y)\sqrt{\mathbf{p}^2-s^2-2\text{i}p_zs}} \end{array}\right),
\end{eqnarray}
\begin{eqnarray} \label{wf2} \nonumber
|\psi_2\rangle&=&\left(\begin{array}{c} -p_x+\text{i}p_y \\ \\ \sqrt{\mathbf{p}^2-s^2+2\text{i}p_zs}+is+p_z  \end{array}\right) \\
|\phi^2\rangle&=&\left(\begin{array}{c}  -\frac{\sqrt{\mathbf{p}^2-s^2-2\text{i}p_zs}+\text{i}s-p_z}{2(p_x+\text{i}p_y)\sqrt{\mathbf{p}^2-s^2-2\text{i}p_zs}}\\ \\ \frac{1}{2\sqrt{\mathbf{p}^2-s^2-2\text{i}p_zs}}
 \end{array}\right).
\end{eqnarray}
The above eigenstates are unique only up to a gauge freedom of $GL(1,\mathbb{C})$ (see Eq.~(\ref{gauge55})).
Any state in 2D Hilbert space $|\Psi\rangle=c^1|\psi_1\rangle+c^2|\psi_2\rangle=c_1|\phi^1\rangle+c_2|\phi^2\rangle$ can be expanded on either the contravariant eigenvectors $|\psi_{1(2)}\rangle$ or covariant ones $|\phi^{1(2)}\rangle$, with the norm being $\langle\Psi|\Psi\rangle=c_1^*c^1+c_2^*c^2$.

According to the Schr\"odinger equation (\ref{Schro}), a state evolves with $t$ as,
\begin{equation}
|\Psi(t)\rangle=c^1\exp(-\frac{\text{i}}{\hbar}E_1t)|\psi_1\rangle+c^2\exp(-\frac{\text{i}}{\hbar}E_2t)|\psi_2\rangle.
\end{equation}
The norm $\langle\Psi(t)|\Psi(t)\rangle$ is not conserved as  $\langle\psi_1|\psi_2\rangle\neq0$.
However, the pseudo-norm $\langle\Psi|X|\Psi\rangle$ is conserved when $p_x^2+p_y^2>s^2$ and $p_z=0$.
With Eqs. (\ref{wf1},\ref{wf2}), we find that
\begin{equation}
X(p_x,p_y)=\frac{1}{2}\left(\begin{array}{cc}1 &-s\frac{p_y+\text{i}p_x}{p_x^2+p_y^2} \\ -s\frac{p_y-\text{i}p_x}{p_x^2+p_y^2}&1 \end{array} \right)\,.
\end{equation}

We turn to the adiabatic evolution and geometric phase. According to our general theory, the adiabatic evolution
is possible only when $E_1$ and $E_2$ are real. This means that for this particular model the adiabatic evolution  can
happen when  $p_x^2+p_y^2>s^2$ and $p_z=0$. When $p_x$ and $p_y$ change slowly on the plane $p_z=0$ with $p_x^2+p_y^2>s$,
an initial eigenstate  $|\psi_{1(2)}(p_{x}(0),p_{y}(0),p_{z}=0)\rangle$ will always be on the instantaneous eigenstate $|\psi_{1(2)}(p_{x}(t),p_{y}(t),p_{z}=0)\rangle$. We study the geometric phase on this plane with constant $s$.

For this example, Eq.~(\ref{conditionY}) does not hold, indicating that the Berry phase is generally complex.
According to Eq.~(\ref{Gcur}), we obtain the purely imaginary  Berry curvature
\begin{equation} \label{Berry-cur3}
 B_{1(2)}=\mp \frac{\text{i}}{2}\frac{s}{(p_x^2+p_y^2-s^2)^\frac{3}{2}},
\end{equation}
with $-/+$ for the state $|\psi_1\rangle/|\psi_2\rangle$. The Berry curvature is in the $p_z$ direction, and is divergent
on the EP ring $p_x^2+p_y^2=s^2$, $p_z=0$. One can check that the Berry curvature
does not change  upon the gauge transformation imposed by $f$ as shown in Eq.~(\ref{gauge55}) whereas the corresponding
Berry connection is modified as shown in Eq.~(\ref{gaugetrans}).  Our general theory dictates that, as $(p_x,p_y)$ change adiabatically around a loop in the plane $p_x^2+p_y^2>s$, $p_z=0$, the state returns to the initial state but with a purely imaginary geometric phase,
\begin{eqnarray}  \nonumber
|\psi\rangle&=&|\psi_1(p_{x}(0),p_{y}(0))\rangle e^{\text{i} \int_{S} \mathbf{B}_{1} dp_x dp_y}, \\
&=&|\psi_1(p_{x}(0),p_{y}(0))\rangle e^{\int_{S} \frac{s}{2(p_x^2+p_y^2-s^2)^\frac{3}{2}} dp_x dp_y},
\end{eqnarray}
with $S$ denoting the area enclosed by the loop $\mathcal{C}$. This imaginary geometric phase can be viewed as  a geometric gain
or loss of particles in dissipative systems described by non-Hermitian Hamiltonian~\cite{GarrisonPLA}.

\subsection{Bogoliubov de Gennes equation}

The second example is the simplest Bogoliubov de Gennes system, which has only two modes. Its Hamiltonian reads
\begin{equation} \label{decomposition}
H=\left(\begin{array}{cc}z&y+{\text i}x \\ -y+{\text i}x&-z\end{array} \right)=\text{i}x\sigma_x+\text{i}y\sigma_y+  z\sigma_z,
\end{equation}
where $x$, $y$ and $z$ are real parameters. This Bogoliubov de Gennes Hamiltonian governs the dynamics of Bosonic Bogoliubov quasiparticles.
Its eigenenergies are
\begin{equation}
E_{1(2)}=\pm\sqrt{z^2-x^2-y^2},
\end{equation}
which are real when $z^2\geq x^2+y^2$. In the parameter space spanned by $(x,y,z)$, all points on
the surface of the cone $z^2=x^2+y^2$ are EPs as one can show that the algebraic multiplicity $\eta(0)=2$ but
 the geometric multiplicity $\zeta(0)=1$ on the degenerate cone. Off the cone, we have $\eta(E_1)=\zeta(E_1)=\eta(E_2)=\zeta(E_2)=1$.

In a certain gauge, the biorthonormal  contravariant and covariant eigenvectors can be worked out as
\begin{eqnarray} \label{gauge11} \nonumber
|\psi_1\rangle&=&\left(\begin{array}{c}a \\ b \end{array} \right), \quad |\phi^1\rangle=\left(\begin{array}{c}a \\ -b \end{array} \right)   \\
|\psi_2\rangle&=&\left(\begin{array}{c}b^* \\ a^* \end{array} \right), \quad |\phi^2\rangle=\left(\begin{array}{c}b^* \\ -a^* \end{array} \right),
\end{eqnarray}
where
\begin{eqnarray} \nonumber
a(x,y,x)&=&-\frac{1}{\sqrt{2}}\frac{z+\sqrt{z^2-x^2-y^2}}{\sqrt{z^2-x^2-y^2+z\sqrt{z^2-x^2-y^2}}},  \\ b(x,y,z)&=&\frac{1}{\sqrt{2}}\frac{y-x\text{i}}{\sqrt{z^2-x^2-y^2+z\sqrt{z^2-x^2-y^2}}}.
\end{eqnarray}
The biorthonormal states can be modified freely by a gauge transformation as shown in Eq.~(\ref{gauge55}).

Under the Bogoliubov de Gennes equation, the norm $\langle\Psi|\Psi\rangle$ of a general state
\begin{equation}
|\Psi\rangle=c^1\exp(-\frac{\text{i}}{\hbar}E_1t)|\psi_1\rangle+c^2\exp(-\frac{\text{i}}{\hbar}E_2t)|\psi_2\rangle
\end{equation}
is not conserved. Instead, what is conserved during the temporal evolution is the pseudo-norm $\langle\Psi|X|\Psi\rangle$ where
\begin{equation}
X(x,y,z)=\left(\begin{array}{cc}|a|^2+|b|^2&-2ab^* \\ -2a^*b&|a|^2+|b|^2 \end{array} \right)\,.
\end{equation}

 The adiabatic evolution can occur when $(x,y,z)$ change slowly inside the cone $z^2>x^2+y^2$ where the eigenenergies are real.
 In this example, we can find that Eq.~(\ref{conditionY}) holds, i.e., $\langle \psi_{1(2)}|\frac{\partial X}{\partial \mathbf{R}}|\psi_{1(2)}\rangle=0$, indicating that the Berry phase becomes real. According to Eq.~(\ref{Gcur}) we find that
 the  Berry curvature is~\cite{ZhangNJP}
\begin{equation} \label{density}
\mathbf{B}=\mp \frac{(1+\tan^2\theta)^{\frac{3}{2}}}{2(1-\tan^2\theta)^{\frac{3}{2}}}\hat{\mathbf{R}},
\end{equation}
with $-/+$ associated with the state $|\psi_1\rangle /|\psi_2\rangle$), $\theta=\text{atan}\left(\frac{\sqrt{x^2+y^2}}{z}\right)$ and $\mathbf{R}\equiv(x,y,x)$ ($\hat{\mathbf{R}}$ is the unit vector along $\mathbf{R}$). Upon the gauge transformation (\ref{gauge55}), the Berry connection is modified according to Eq.~(\ref{gaugetrans}) but  the Berry curvature is fixed. The Berry curvature becomes divergent as $\theta\rightarrow\pm\pi/4$, i.e., on the degenerate cone determined by $z=\pm\sqrt{x^2+y^2}$, indicating that these EPs on the cone are  monopoles.

In this example, the constant $Y$ matrix as shown in Eq.~(\ref{relation21}) exists and it is just one of the Pauli matricess
\begin{equation}
Y=\sigma_z=\left(\begin{array}{cc} 1&0\\0&-1 \end{array}\right),
\end{equation}
with $\alpha_1=1$ and $\alpha_2=-1$ defined in Eq.~(\ref{relation25}). We then have the simple relations $|\phi^{1}\rangle= Y|\psi_{1}\rangle$, $|\phi^{2}\rangle=-Y|\psi_{2}\rangle$ and the Berry connections according to Eq.~(\ref{Berry-conn2}) ~\cite{ZhangNiu,ZhangNJP},
\begin{equation}
\mathbf{A}_{1(2)}=\pm\text{i}\langle\psi_{1(2)}\sigma_z\frac{\partial}{\partial \mathbf{R}}|\psi_{1(2)}\rangle,
\end{equation}
with $+/-$ associated with the state $|\psi_1\rangle /|\psi_2\rangle$). Geometric phase being real indicates that
there is no geometric  gain or loss of the particles during the adiabatic evolution.

The Chern number, which reflects the total magnetic charge contained by the monopole, can be calculated from Eq.~(\ref{density}) as,
\begin{equation}
\mathcal{C}_n\rightarrow\mp\infty\,.
\end{equation}
This is drastically different from the  Chern numbers in Hermitian systems, which are always $2n\pi$ with $n$ being integer.

\section{Summary}

To summarize, we have studied the adiabatic geometric phase of non-Hermitian quantum mechanics. We show that the structure of geometric phase of non-Hermitian quantum mechanics is quite different from the unitary quantum mechanics. Since such non-Hermitian dynamics can be generically found or constructed in various physical systems,  our results provide   new insights into these non-Hermitian systems.
The present work also provides a new perspective toward the fundamental understanding of quantum evolution.

\section{acknowledgement}
This work was supported by the The National Key Research and Development Program of China (Grants No.~2017YFA0303302, No.~2018YFA030562)
and the National Natural Science Foundation of China (Grants No.~11334001 and No.~11429402).


\begin{thebibliography}{99}
\bibitem{ZhangNJP} Qi Zhang and Biao Wu, New. J. Phys. {\bf 20} 013024 (2018).
\bibitem{njp} B. Wu and Q. Niu, New J. of Phys. {\bf 5}, 104 (2003).

\bibitem{Bender} C.M. Bender and S. Boettcher, Phys. Rev. Lett. {\bf 80}, 5243 (1998).
\bibitem{Bender2} C.M. Bender, D.C. Brody and H.F. Jones, Phys. Rev. Lett. {\bf 89}, 270401 (2002).
\bibitem{Bender3} C.M. Bender, Rep. Prog. Phys. {\bf 70} 947 (2007).
\bibitem{Mostafazadeh} A. Mostafazadeh, J. Math. Phys. {\bf 43}, 205 (2002); {\bf 43}, 2814 (2002).
\bibitem{Berry} M. V. Berry, J. Opt. {\bf 13}, 115701 (2011).
\bibitem{Longhi} S. Longhi, Phys. Rev. Lett. {\bf 103}, 123601 (2009).
\bibitem{West} C.T.West, T.Kottos, and T. Prosen,Phys. Rev. Lett. {\bf 104}, 054102 (2010).
\bibitem{Mostafazadeh2} A. Mostafazadeh, J. Math. Phys. {\bf 44}, 974 (2003).
\bibitem{Bender4} C.M. Bender, P.N. Meisinger, and Q. Wang, J. Phys. A: Math. Gen. {\bf 36}, 1029 (2003).

\bibitem{Bender5} C.M. Bender, P.N. Meisinger, and Q. Wang, J. Phys. A: Math. Gen. {\bf 36} 6791 (2003).


\bibitem{Mostafazadeh3} A. Mostafazadeh, J. Phys. A: Math. Gen. {\bf 36}, 7081 (2003).

\bibitem{Mostafazadeh4} A. Mostafazadeh A and S. \"Ozcelik, Turk. J. Phys. {\bf 30}, 437  (2006).

\bibitem{Wang} Q. Wang, J. Phys. A: Math. Gen. {\bf 43}, 295301 (2010).

\bibitem{chen} Liping Guo, Lei Du, Chuanhao Yin, Yunbo Zhang, and Shu Chen, Phys. Rev. A {\bf 97}, 032109 (2018).

\bibitem{e1} Z. H. Musslimani, K. G. Makris, R. El-Ganainy, and D. N.
Christodoulides, Phys. Rev. Lett. {\bf 100}, 030402 (2008).
\bibitem{e2} K. G. Makris, R. El-Ganainy, D. N. Christodoulides, and Z. H.
Musslimani, Phys. Rev. Lett. {\bf 100}, 103904 (2008).
\bibitem{e3} A. Guo, G. J. Salamo, D. Duchesne, R.Morandotti, M. Volatier-
Ravat, V. Aimez, G. A. Siviloglou, and D. N. Christodoulides,
Phys. Rev. Lett. {\bf 103}, 093902 (2009).
\bibitem{e4} C. E. Ruter, K.G.Makris, R. El-Ganainy, D.N. Christodoulides,
M. Segev, and D. Kip, Nat. Phys. {\bf 6}, 192 (2010).
\bibitem{e5} J. Schindler,  A. Li,  M. C. Zheng,  F. M. Ellis, and T. Kottos,
Phys. Rev. A {\bf 84}, 040101 (2011).
\bibitem{e6} A. Regensburger, C. Bersch, M. A. Miri, G. Onishchukov, D.
N. Christodoulides, and U. Peschel, Nature (London) {\bf 488}, 167
(2012).
\bibitem{e7} A. Regensburger, M. A. Miri, C. Bersch, J. Nager, G. Onishchukov,
D. N. Christodoulides, and U. Peschel, Phys. Rev.
Lett. {\bf 110}, 223902 (2013).
\bibitem{e8} M. Brandstetter, M. Liertzer, C. Deutsch, P. Klang, J. Schoberl,
H. E. Tureci, G. Strasser, K. Unterrainer, and S. Rotter,
Nat. Commun. {\bf 5}, 4034 (2014).
\bibitem{e9} B. Peng, S. K. Ozdemir, S. Rotter, H. Yilmaz, M. Liertzer, F.
Monifi, C. M. Bender, F. Nori, and L. Yang, Science {\bf 346}, 328
(2014).
\bibitem{e10} B. Peng, S. K. Ozdemir, F. Lei, F. Monifi, M. Gianfreda, G. L.
Long, S. Fan, F. Nori, C. M. Bender, and L. Yang, Nat. Phys.
{\bf 10},  394 (2014).
\bibitem{e11} H. Jing, S. K. Ozdemir, X.-Y. Lu, J. Zhang, L. Yang, and F. Nori,
Phys. Rev. Lett. {\bf 113}, 053604 (2014).
\bibitem{e12} L. Feng, Z. J. Wong, R.-M. Ma, Y. Wang, and X. Zhang,
Science {\bf 346}, 972 (2014).
\bibitem{e13} M. Liertzer, L. Ge, A. Cerjan, A. D. Stone, H. E. Tureci, and S.
Rotter, Phys. Rev. Lett. {\bf 108}, 173901 (2012).

\bibitem{J1} Zhong-Peng Liu {\it et.al.}, Phys. Rev. Lett. {\bf 117}, 110802 (2016)
\bibitem{J3} Xin-You L\"u, Hui Jing, Jin-Yong Ma, and Ying Wu, Phys. Rev. Lett. {\bf 114}, 253601 (2015).
\bibitem{J4} H. Jing {\it et.al.},  Sci. Rep. {\bf 5}, 9663 (2015).
\bibitem{J5} Jing Zhang {\it et.al.}, Phys. Rev. B {\bf 92}, 115407 (2015).

\bibitem{T1} K. Esaki, M. Sato, K. Hasebe, and M. Kohmoto, Phys.
Rev. B {\bf 84}, 205128 (2011).
\bibitem{T2} S.-D. Liang and G.-Y. Huang, Phys. Rev. A {\bf 87}, 012118
(2013).
\bibitem{T3} T. E. Lee, Phys. Rev. Lett. {\bf 116}, 133903 (2016).
\bibitem{T4} D. Leykam, K. Y. Bliokh, C. Huang, Y. D. Chong, and
F. Nori, Phys. Rev. Lett. {\bf 118}, 040401 (2017).
\bibitem{T5} H. Menke and M. M. Hirschmann, Phys. Rev. B {\bf 95},
174506 (2017).
\bibitem{T6} Y. Xu, S.-T. Wang, and L.-M. Duan, Phys. Rev. Lett.
{\bf 118}, 045701 (2017).
\bibitem{T7} J. Gonzsalez and R. A. Molina, Phys. Rev. B {\bf 96}, 045437
(2017).
\bibitem{T8} W. Hu, H. Wang, P. P. Shum, and Y. D. Chong, Phys.
Rev. B {\bf 95}, 184306 (2017).
\bibitem{T9} Y. Xiong, arXiv:1705.06039.
\bibitem{Fu} Huitao Shen, Bo Zhen, Liang Fu, Phys. Rev. Lett. {\bf 120}, 146402 (2018).

\bibitem{chen1} Chuanhao Yin, Hui Jiang, Linhu Li, Rong L\"u, and Shu Chen, Phys. Rev. A {\bf 97}, 052115 (2018).
\bibitem{chen2} Hui Jiang, Chao Yang, and Shu Chen, arXiv:1809.00850.


\bibitem{Nering} Evar D. Nering, Linear Algebra and Matrix Theory (2nd ed.), New York: Wiley, LCCN 76091646 (1970).
\bibitem{F3} A. Mostafazadeh, J. Math. Phys. {\bf 43}, 3944 (2002).
\bibitem{F1} A. Mostafazadeh, J. Math. Phys. {\bf 43}, 205 (2002).
\bibitem{F2} A. Mostafazadeh, J. Math. Phys. {\bf 43}, 2814 (2002).
\bibitem{GarrisonPLA} J. G. Garrison and E. M. Wright, Phys. Lett. A {\bf 128}, 177 (1988)
\bibitem{GePRA} Y. C. Ge and M. S. Child, Phys. Rev. A {\bf 58}, 872 (1998).


\bibitem{PLA} B. Wu, Q. Zhang, and J. Liu, Phys. Lett. A {\bf 375}, 545 (2011).











\bibitem{ZhangNiu} C. Zhang, A. M. Dudarev, and Q. Niu, Phys. Rev. Lett. {\bf 97}, 040401 (2006).

\end{thebibliography}
\end{document}